Title
- Fabrication of Poly (ε-Caprolactone) 3D scaffolds with controllable porosity using ultrasound
- 3D Printing of Polycaprolactone with ultrasound

Authors

Martin Weber,[1]* Dmitry Nikolaev,[1] Mikko Koskenniemi,[1] Jere Hyvönen,[1] Joel Jääskeläinen,[1] Armand Navarre,[2] Ekaterina Takmakova,[3] Arun Teotia,[4] Pekka Katajisto,[4] Robert Luxenhofer,[3] Edward Hæggström[1], Ari Salmi[1]

Affiliations

* martin.weber@helsinki.fi

[1] Electronics Research Laboratory, Department of Physics, Faculty of Science, University of Helsinki, Helsinki, Finland.

[2] Technical college of Blois, Blois, France.

[3] Soft Matter Chemistry, Department of Chemistry, and Helsinki Institute of Sustainability Science, Faculty of Science, University of Helsinki, Helsinki, Finland.

[4] Institute of Biotechnology (HiLIFE), Faculty of Biological and Environmental Sciences, University of Helsinki, Helsinki, Finland.

Abstract

3D printing has progressed significantly, allowing objects to be produced using a wide variety of materials. Recent advances have employed focused ultrasound in 3D printing, to allow printing inside acoustically transparent materials. Here we introduce a Selective Ultrasonic Melting (SUM) method for 3D printing of poly (ε-caprolactone) (PCL) powder mixed with water. The printing was done by mechanically moving a focused ultrasound transducer. The microstructure and porosity of the prints were analyzed with micro-computed tomography (μCT). The open porosity of the printed samples was determined using the water intrusion method and by passing fluorescent microspheres through the structure. The cytocompatibility of the printed structures was confirmed by seeding NIH-3T3 fibroblast cells on the scaffolds, followed by analysis using live/dead fluorescent assay. and visualization using scanning electron microscopy (SEM). We demonstrated that SUM is a viable technique to print structures with active control of their porosity This method provides an alternative to methods such as fused deposition modelling (FDM) and material jetting.

Teaser

3D objects with controllable porosity were printed from polycaprolactone powder using ultrasonic energy.

## MAIN TEXT
## Introduction

Additive manufacturing uses a variety of techniques to construct objects by melting powders composed of submillimeter particles, such as selective laser sintering (SLS), selective laser melting (SLM), and electron beam melting (EBM) (*1*). These methods use a focused laser or electron beam to heat the material and fuse the powder particles. Since neither the laser nor the electron beams can penetrate deeply into the powder, the printing is performed layer-by-layer with additional powder applied between the processing of each layer. Another emerging approach involves the use of focused ultrasound (FUS), which can penetrate a wide range of materials and concentrate acoustic energy precisely within a desired region.

FUS can rapidly cure silicone polymers via sonochemical reactions (*2*), and holographic sound printing enables cross-sectional polymerisation of silicone elastomers (*3*). FUS can initiate polymerization reactions of self-enhancing sonicated ink within deep layers of tissue via sonothermal-induced gelation (*4, 5*). A mixture of natural products such as egg white and potato flour can be sonothermally solidified and used to print 3D objects (*6*). Acoustic radiation force can create 3D objects through particle assembly and material structuring which includes the use of acoustic holograms in liquid (*7*) and acoustic field control with phased arrays in air (*8*).

3D printing promises use in personalized medical applications, from implants and prosthetics to tissueengineered structures for tissue regeneration (*9, 10*). Notably, porous structures that mimic extracellular matrices are crucial in bone tissue engineering (*11-14*) and in drug delivery systems (*15, 16*). PCL is widely used in these applications, offering advantages such as biodegradability, processability, biocompatibility, and robust mechanical properties (*15, 17-19*).

Due to the thermoplastic properties of PCL, the most common printing technique is fused deposition modelling (FDM) (*20, 21*) which allows printing cell-loaded scaffolds with controlled porosity (*22*). This method suffers from low resolution, manufacturing complexity, and unsuitability for in situ printing.

We introduce the Selective Ultrasonic Melting (SUM) method that uses FUS to fuse PCL powder mixed with water into solid structures. In this technique, the ultrasonic beam penetrates the PCL sample, releasing thermal energy in the focus, which causes the PCL to melt, forming a voxel. By moving the focal spot within the sample, a porous 3D structure is formed.

We used the NIH-3T3 fibroblast cell line to verify the cytocompatibility of structures printed at different feed rates. Cell viability was determined using live/dead fluorescent imaging. Furthermore, the cell-laden scaffolds were visualized by SEM imaging.

## Results

### Printed samples

We printed structurally stable samples with repeatable print quality, achieving a maximum thickness of 5 mm. The printed ETLA logo at different feed rates as well as the samples for the cell tests, all produced at a constant transducer power, are shown in **Figure *1***. Each sample had a thickness of 3 mm and the printing took 5 to 10 minutes. All printed objects remained mechanically stable and intact after drying. The samples printed at different feed rates exhibited pronounced differences in mechanical properties: e.g. the ETLA logo printed at 100 mm min$^{-1}$ had a solid core and was rigid, whereas the 200 mm min$^{-1}$ print was brittle and highly porous.

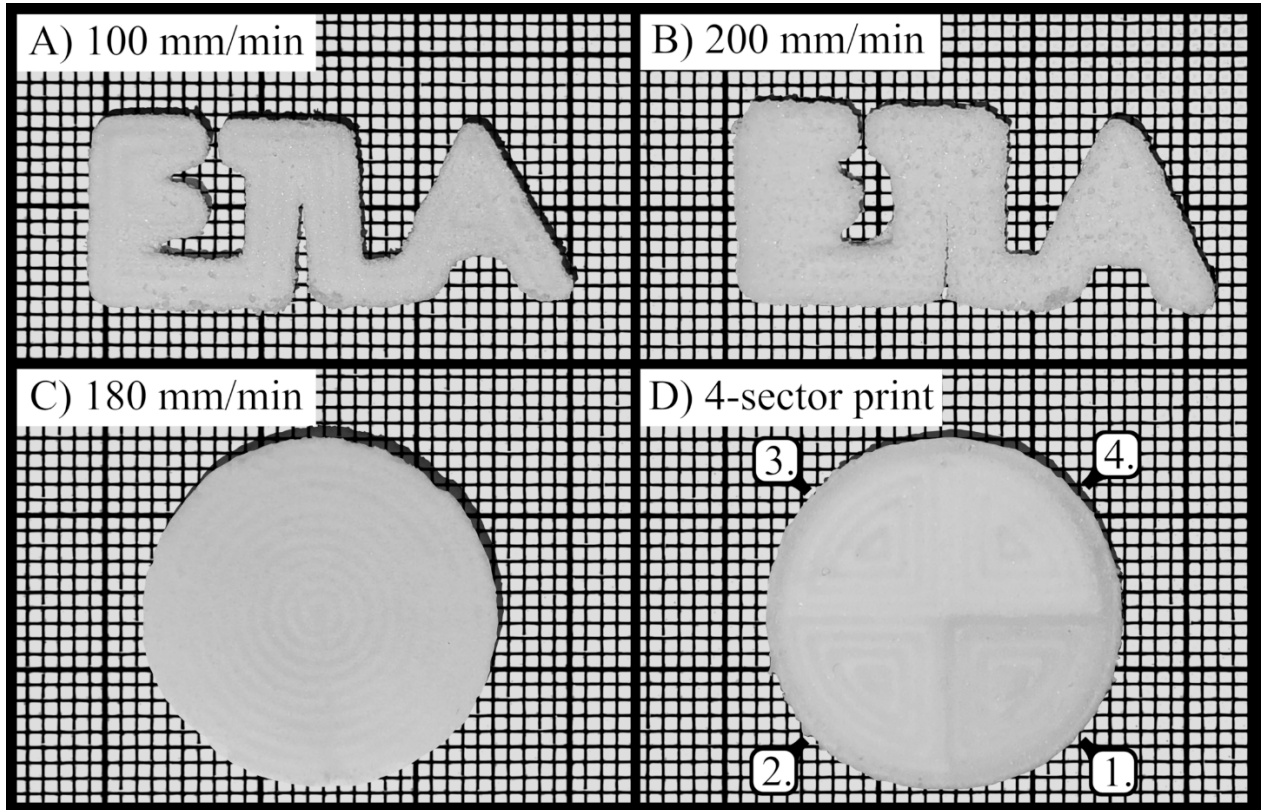

**Figure 1: Printed samples.** ETLA logos printed at a feed rate of 100 mm min$^{-1}$ (A) and 200 mm min$^{-1}$ (B) are shown in the top row. A disk with a single speed of 180 mm min$^{-1}$ (C) and a four sector disk (D) are shown in the bottom row. Speeds on the four sector disk are labeled from 1 to 4 with speeds of 25 mm min$^{-1}$, 75 mm min$^{-1}$, 125 mm min$^{-1}$ and 170 mm min$^{-1}$ respectively. The bold grid is 10 mm

### Simulation of the heat release in the focus

The peak acoustic power calculated from the measured hydrophone scan was 53.6 W, with an average acoustic power of 1.03 W. The measured attenuation coefficient and sound speed of the PCL-water suspension at a frequency of 4.2 MHz were 2.0 dB cm$^{-1}$ and 1760 m s$^{-1}$, respectively. The peak acoustic power and acoustic properties of the PCL-water suspension were used to model an equivalent piston focusing source in the *HIFU beam* software for simulating the nonlinear propagation of a radially symmetric ultrasound beam in a layered medium composed of absorbing materials.

The geometry of the problem is shown in **Figure 2** A. The medium consisted of two layers: water and the PCL-water suspension. The geometric focus of the transducer was located 2 mm from the interface in the suspension layer.

The simulation showed that the acoustic intensity at the focus reached 2.3 kW/cm$^2$ with a corresponding negative peak pressure of 17.4 MPa (mechanical index of 8.5 MPa MHz$^{-0.5}$). The heat release distribution (**Figure 2** B) has a maximum value of 63 kW cm$^{-3}$. At this level of heat power density, PCL begins to melt after 90 ms of sonication. A sample produced by sonicating the material at one point for 5 s is shown in the corner of **Figure 2** B.

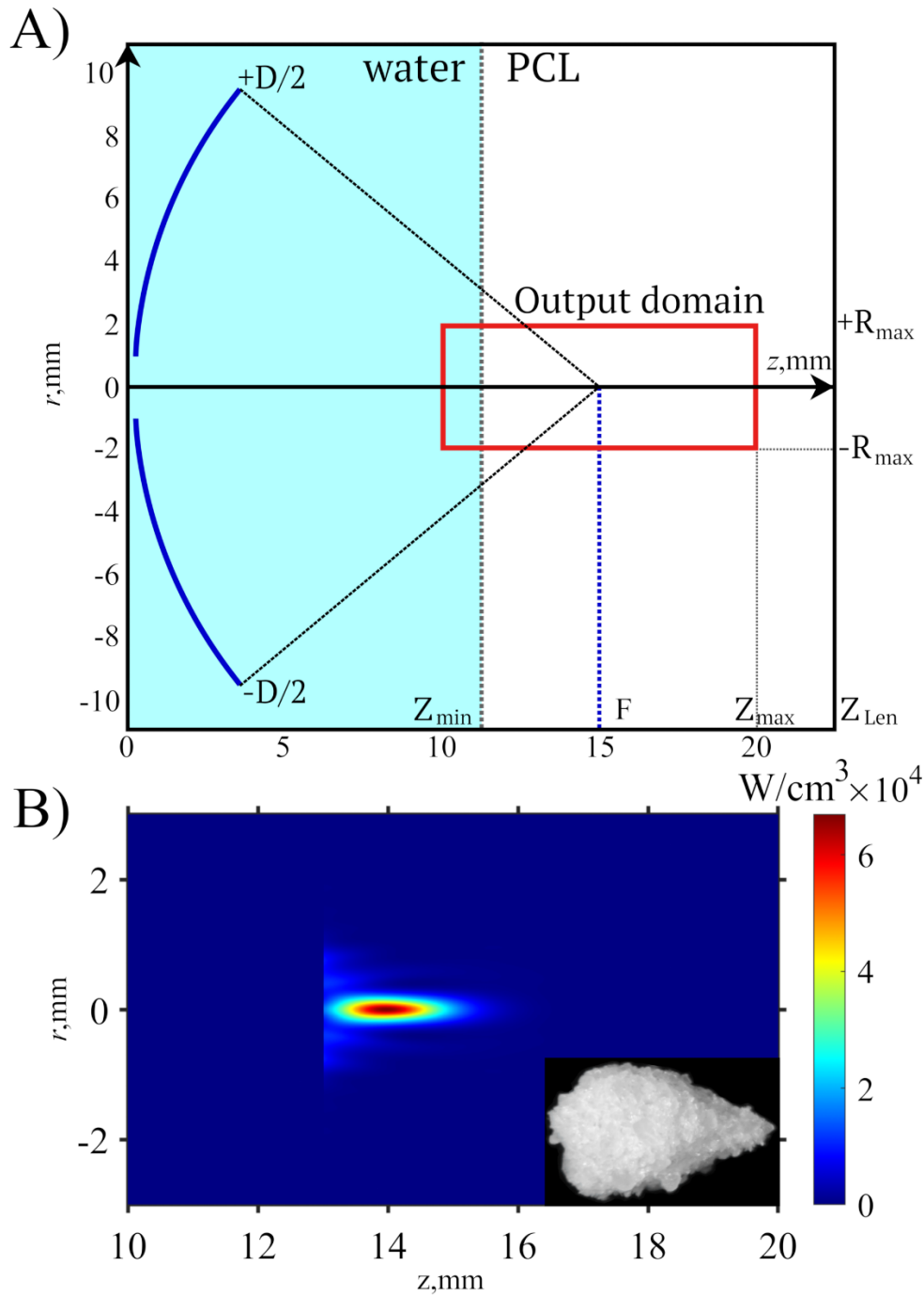

**Figure 2: Acoustic field simulation.** Geometry of the acoustic field simulation (A) and the simulated heat deposition in the focal area of the output domain (B). A scale matched five-second-sonicated point of PCL is shown in comparison to the focal area.

## Microstructure of the printed samples

**Figure 3** shows μCT images of the microstructure in the center of the samples printed at different feed rates. At feed rates up to 100 mm min$^{-1}$, the powder particles fused, forming large, interconnected structures. In contrast, at feed rates of 125 mm min$^{-1}$ and above, the material's microstructure resembles that of the untreated reference sample.

**Figure 4** shows the PCL porosity in the samples calculated from the μCT 3D scans. The decreasing porosity of PCL at low feed rates indicates that the powder is completely melted, leading to grain fusion and displacement of water from the samples. At high feed rates, the porosity approaches that of the untreated sample, indicating that the powder grains are connected only by tiny bridges formed by surface melting (not resolved in the µCT images). The pores form a net-like structure with a characteristic transverse size of 10 µm.

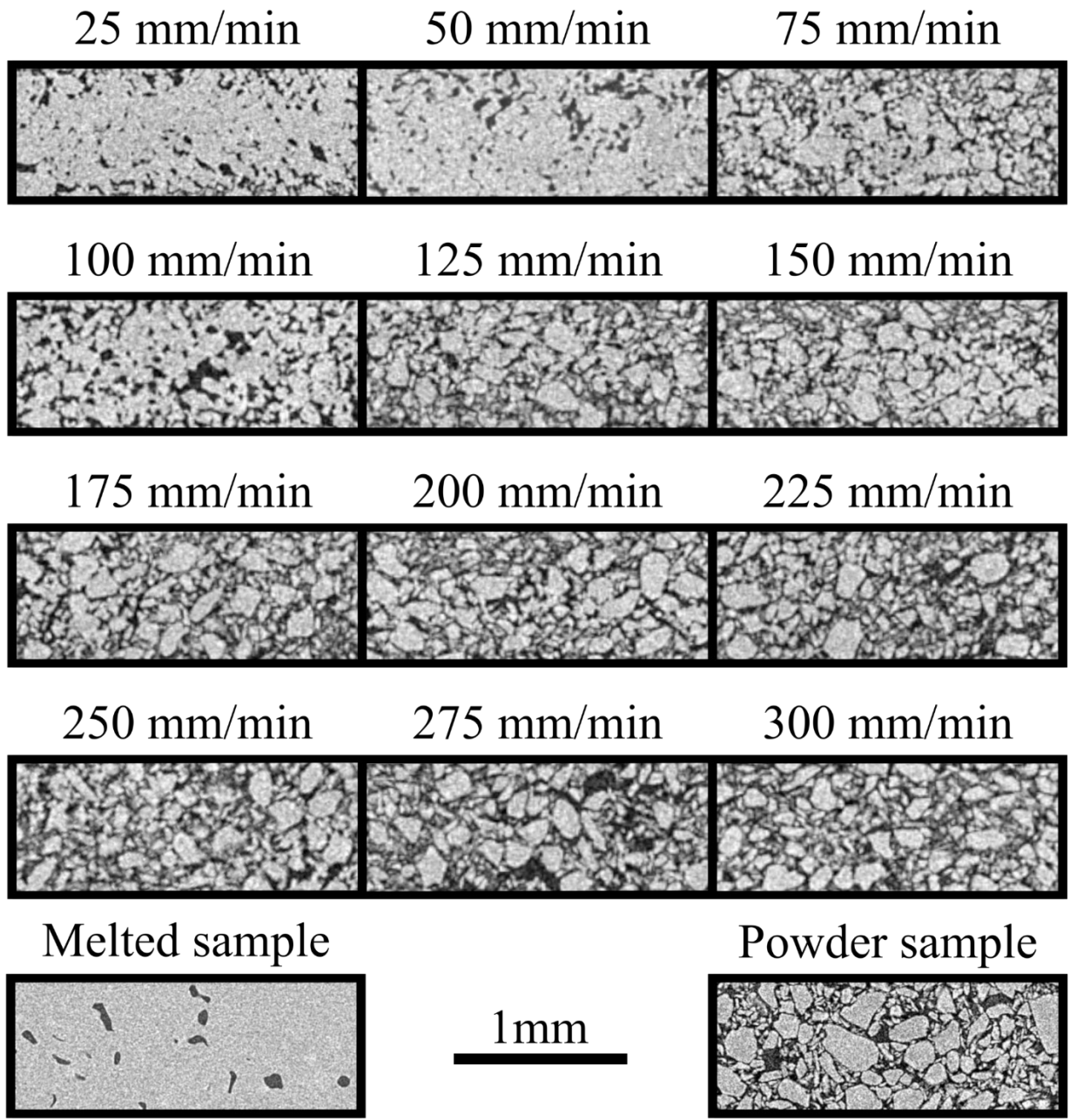

**Figure 3: µCT-imaged microstructure in printed PCL as a function of feed rate:** 25 mm min$^{-1}$ to 300 mm min$^{-1}$. The structure of a thermally melted sample obtained by immersion in boiling water and an unsonicated powder sample are presented for reference. The voxel resolution of the µCT scans is 15 µm for the different feed rates and 7.5 µm for the reference (melted and powder) samples.

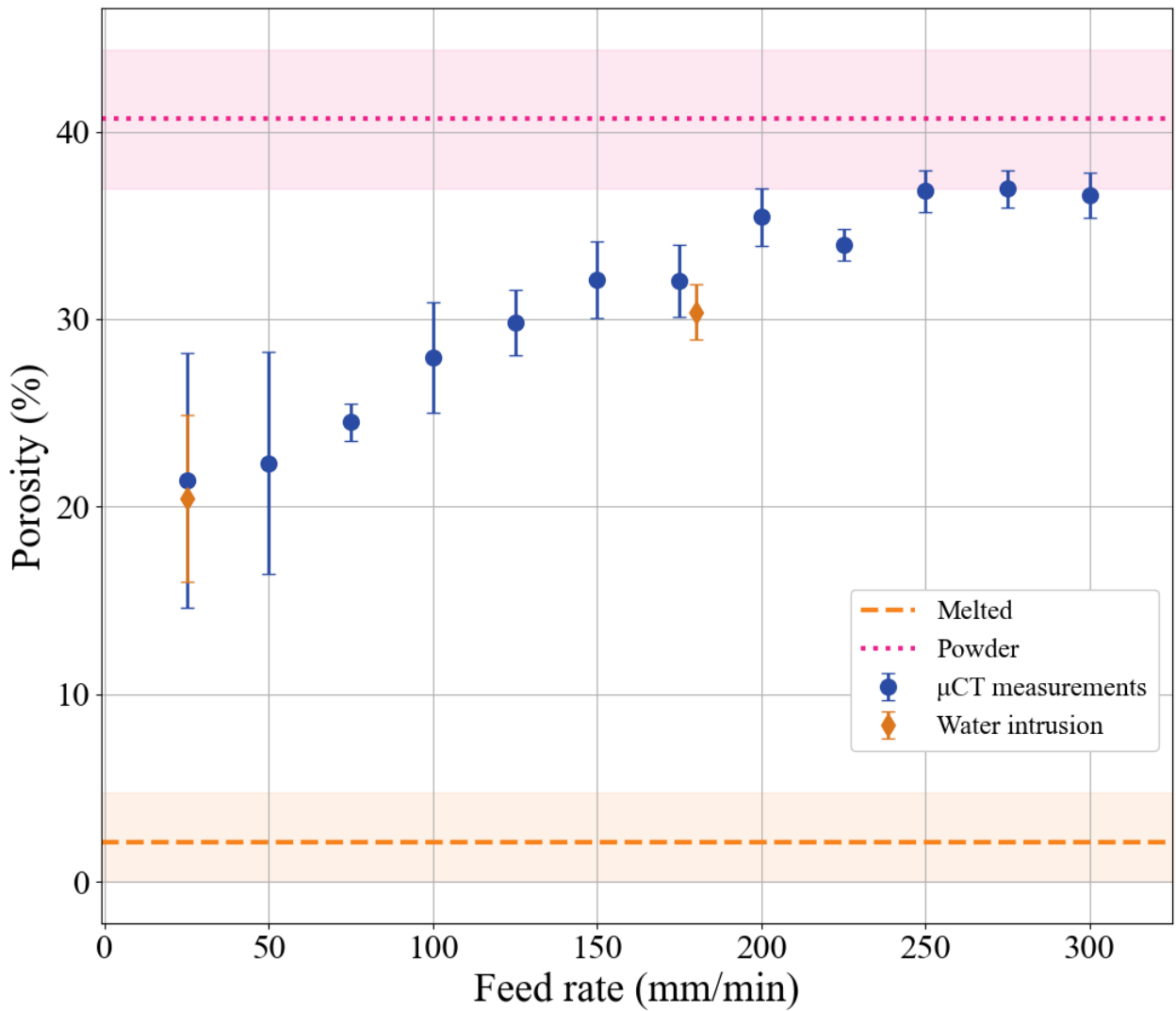

**Figure 4: Porosity of PCL as a function of feed rate.** The porosity of thermally melted PCL and dry powder are shown for reference. For each porosity value, the mean and one standard deviation are shown. Two additional data points from the water intrusion method are shown.

### Interconnection of pores

The interconnection of pores should enable nutrient exchange and potentially cell penetration into the printed scaffold, which is important for tissue engineering applications. Moreover, the pore size determines if cells can populate the scaffold. The PCL volume porosity measurements for the feed rate of 25 mm min$^{-1}$ and 180 mm min$^{-1}$ obtained by the water intrusion method are consistent with those obtained by the µCT method, **Figure 4**. The agreement between the results obtained with both methods indicates that most pores in the samples are interconnected.
**Figure 5** shows sample cross-sections after passing water-containing microspheres through them. The 1 µm microspheres are observed throughout the entire depth of the sample, whereas 10 µm microspheres exhibit limited penetration, mainly accumulating on the surface and in the upper layers of the sample. Water containing both 1 µm and 10 µm spheres was visualized by microscope after passing through the sample (**Figure 5**). The results confirm that the 1 µm spheres readily penetrate through the sample, whereas the 10 µm spheres are mainly trapped inside or on the surface of the sample.

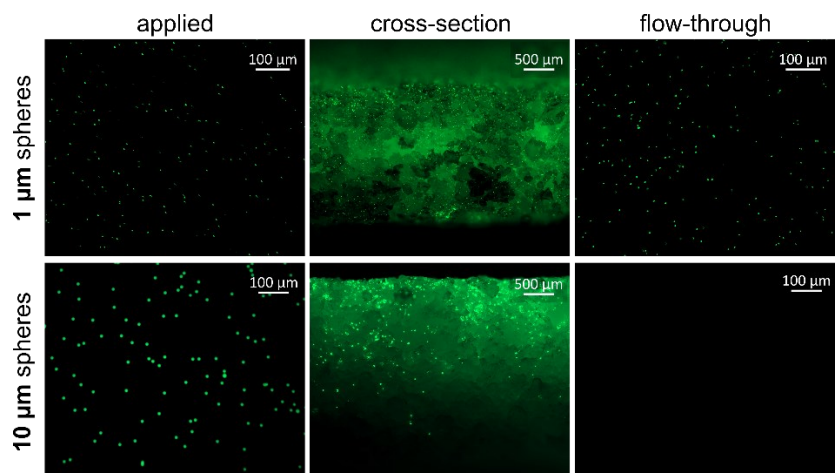

**Figure 5: Microscope images of the distribution of microspheres.** Images of the applied suspension containing fluorescent microspheres, the scaffold crosssections after applying the suspension, and the suspension that flowed through the sample are shown. The samples printed at a feed rate of 300 mm min$^{-1}$ had an average thickness of 3 mm. Suspensions containing microspheres with diameters of 1 μm and 10 μm were applied on the top of the samples. The suspension permeated through the sample, carrying the microspheres into their internal structure.

### Cytocompatibility of the PCL scaffolds printed using SUM

PCL is commonly used in medical applications, hence the cytocompatibility of the printed objects was investigated. To understand whether the microstructure and porosity of the objects affect cell adherence and viability, the four-sector scaffold design (with 25, 75, 125, and 175 mm min$^{-1}$ feed rates) was used (**Figure** *1*D). Live/dead imaging by FDA/PI of fibroblasts cultured on these scaffolds showed >90 % cell viability in all sectors. Representative images of 25 and 175 mm min$^{-1}$ sectors are shown in **Figure** *6*. More examples are presented in supplementary *Figure S1*. Interestingly, the cells were equally distributed among the sectors, suggesting no preference for growing in a specific area.

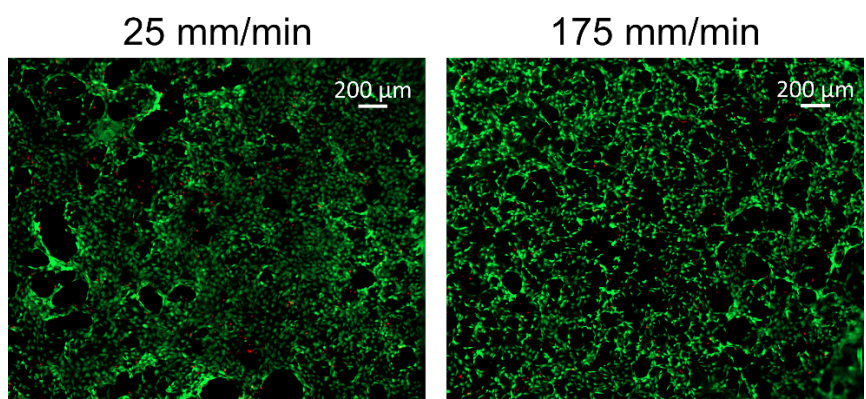

**Figure 6: Cell viability and distribution on the scaffold with four-sector design.** The sectors were printed with 25, 75, 125, and 175 mm min$^{-1}$ feed rates. NIH-3T3 fibroblasts were cultured on this scaffold for three days, followed by the live (green) / dead (red) staining.

### SEM images of cell-laden scaffolds

SEM analysis of the 3D scaffolds provides high-level information about the surface topography of the scaffolds printed at different feed rates. To analyze the surface properties, SEM analysis was done on the printed 3D scaffolds without cell seeding. The fusion of the PCL particles decreased with higher feed rates, leading to increased surface roughness (**Figure 7**(i)). This translates into higher porosity of the scaffolds printed at high feed rates. In contrast, scaffolds printed at low feed rates (**Figure 7**) A(i) and B(i) exhibited a higher degree of particle fusion and smoother surfaces. The SEM analysis of cell seeded scaffolds (**Figure 7**(ii) and (iii)) demonstrated similar results. It was observed that at 25 mm min$^{-1}$ and 75 mm min$^{-1}$ feed rate the scaffolds have smoother surface, where the cells grow in a monolayer fashion with flat extended morphology. In contrast, the cells growing on the high feed rate scaffolds i.e. 125 mm min$^{-1}$ (**Figure 7**C(ii, iii)) and 175 mm min$^{-1}$ (**Figure 7**D(ii, iii)), demonstrate more 3D-morphology with anchoring filopodia. Formation of such structures by cells indicates cell proliferation on the scaffolds. Although PCL itself is hydrophobic and hence do not feature cell anchoring motifs, serum treatment of the scaffolds might lead to surface protein adsorption enabling cellular adhesion and proliferation on the printed scaffolds. Similar results were observed with NaOH treated scaffolds (*Figure S4*). Here prominent cell adhesion to the scaffold surface, either as monolayers or with 3D extended morphology were observed. In both cases, the results show cell adhesion and proliferation with no significant difference in the cytocompatibility of the scaffolds.

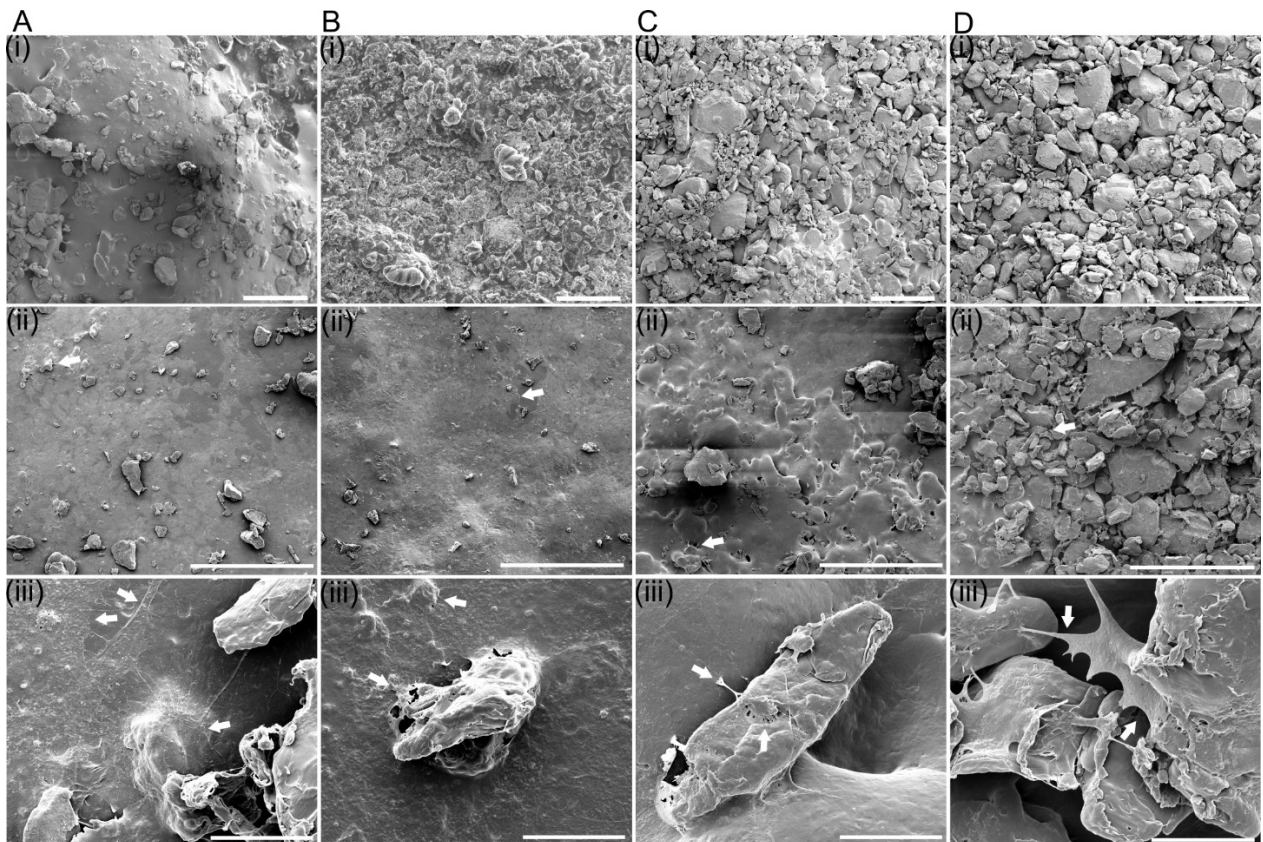

**Figure 7: SEM images of the PCL scaffolds.** Images showing surface topography of 3D scaffolds printed at different feed rates. (A) 25 mm min$^{-1}$, (B) 75 mm min$^{-1}$, (C) 125 mm min$^{-1}$, and (D) 175 mm min$^{-1}$. (i) Images showing control scaffolds not seeded with cells, (ii) and (iii) showing images of cells (white arrows) growing on the respective scaffold surface, either as monolayers or with 3D morphology. Scale bar- (i),(ii)500 µm, (iii)20 µm.

## Discussion

Printing PCL samples using SUM offers unique advantages. Due to its biodegradability and slow degradation rate, PCL is valued in biomedical applications, including tissue engineering, regenerative medicine, and drug delivery (*23*).

Page **8** of 19

The adjustment of porosity provides an effective way to control the degradation rate of PCL, with higher porosity resulting in accelerated weight loss of the material (*11*). Furthermore, controlling the porosity and connectivity of particles allows tuning the mechanical properties of the scaffold, while optimizing the surface roughness improves its biological performance (*24*).

Further customization of the structural, mechanical, surface, and biocompatible properties of the samples can be achieved by incorporating additional components into the base material and by selecting different liquids for the suspension preparation. The bulk mechanical properties of the printed samples can be tailored by adding other materials that are blendable with PCL such as polymers, starch, lignin, etc. (*25, 26*). For example, incorporating silica aerogel into PCL results in modified properties suitable for use in scaffolds for artificial bone (*27, 28*). Adjusting the particle size and shape of the powder grains can lead to changes in structural characteristics, such as increased porosity or larger pore size. Additionally, replacing water with other aqueous substances enables further tuning of material properties. For example, embedding nutrients directly into the material can promote cell growth.

PCL-based material used for SUM printing offers advantages over silicone rubber (*2*). Silicone rubber requires mechanical separation of the printed object from the soft, sticky uncured material, such rubber has a limited pot life that constrains print time, and generates waste since it can not be reused. In contrast, PCL avoids these limitations since it is stable and can be reused.

Resolution is a key factor determining print quality, which is mainly affected by the power and frequency of the acoustic waves, the printer's feed rate, and the suspension's acoustic properties. Both feed rate and acoustic power affect the amount of acoustic energy per mass of material focused into the printing area. This energy is eventually converted into heat, which diffuses and reduces print resolution as more energy is delivered. Increasing the frequency can improve resolution, since high frequencies create a small focal size. However, high frequencies increase the acoustic attenuation which limits penetration depth and restricts the maximum thickness of printed objects. When selecting the operating frequency the wavelength in the material should be significantly larger than the grain size in the suspension to prevent scattering based attenuation (*29*). In the current study the maximum thickness of the printed samples was 5 mm. Thus using suspensions with smaller grain size can improve the resolution. For maximum resolution, the acoustic absorption in the material should be as small as possible, allowing the material to be heated primarily by nonlinear effects. Focusing high intensity ultrasound into very small volumes allows for localized and effective heating (*30*). Therefore, an optimal approach to enhancing print resolution involves minimizing attenuation by employing low frequencies, reducing the grain size of the powder, and by applying short, high amplitude bursts to achieve highly localized heat deposition.

The patterns of the prints (**Figure *1***) with visible printing paths resemble the patterns of objects produced using FDM printing. Similarly to FDM printing, the mechanical strength of ultrasonically printed samples is influenced by the slicing parameters, particularly the spacing between the printing paths. The line separation distance determines the degree of fusion between adjacent paths in both horizontal and vertical directions. Similarly to FDM printing, where the extruded filament thickness is defined by the translation speed and extrusion rate; in ultrasonic printing, the molten path thickness is determined by the feed rate and acoustic power. **Figure *1***C illustrates the variation in the thickness of the printing path for different feed rates at constant acoustic power. At the lowest feed rate, the molten path (darker regions) is thick and nearly fused with the adjacent paths, though a small amount of partially fused powder (lighter regions) remains between the paths. At high feed rates, the molten path becomes thinner and the presence of unmelted powder between the printing paths becomes more pronounced. Unmelted parts reduce the overall structural integrity of the sample. Thus, unlike FDM printing, an ultrasonically printed sample consists not only of molten paths but also includes partially fused material connected to these paths.

The experiments with the fluorescent microspheres confirmed that the pores are interconnected and that particles with at least a size of 1 μm can permeate in the printed object. The 10 μm microspheres were unable to penetrate through the print done at 300 mm min$^{-1}$. Thus, the typical pore size in unfused samples (printed at high feed rates) is between 1 μm and 10 μm. The pore size would be expected to be smaller with slow feed rates when considering that the porosity of PCL increases with increasing feed rates as seen in **Figure 4**.

NIH-3T3 fibroblasts seeded on the printed PCL scaffolds were attached and proliferated, which indicates that the scaffolds are cytocompatible (**Figure 6**, **Figure 7**, *Figure S1* and *Figure S4*). Of note, NIH-3T3 cells can be maintained under standard cell culture conditions using different serums, specifically, calf serum or fetal bovine serum (FBS) (*31, 32*). However, in our experiments the serum choice turned out to be crucial for the survival of these cells in case they were seeded on the PCL scaffolds (*Figure S2*). NIH-3T3 cell viability remained high when they were cultured with the calf serum, but not with the FBS. Interestingly, coating the PCL surface could improve both cell attachment and proliferation on it. For example, poly-D-lysine and collagen type-I coating were used previously (*22*). Taking into account that calf serum and FBS have different compositions, it remains to be investigated which components of calf serum improved the NIH-3T3 cell attachment.

Another PCL surface modification method is sodium hydroxide (NaOH) etching, which increases the hydrophilicity of the PCL scaffold, and thus, cell attachment to it (*33*). However, treatment of the ultrasound printed scaffolds with 5M NaOH for 30 minutes did not influence the ability of NIH-3T3 fibroblasts to grow on those scaffolds (*Figure S3*).

The high-resolution SEM imaging of the 3D printed scaffolds demonstrated a correlation between feed rate and the porosity of the printed scaffolds. Slow feed rates demonstrated high PCL particle fusion and material consolidation, leading to a smooth printed surface. This leads to low structural porosity of the printed scaffolds. Such low porosity structures may provide high mechanical strength to the structure but the lack of macro-porosity hinders cell penetration into the structure where they were mostly growing as monolayers on the surface. Alternatively, high printing feed rates lead to low particle fusion with much higher surface roughness and structural porosity in the printed scaffold. With the SEM analysis cells were observed to grow in a more 3D morphology on these scaffolds with formations of extended cell adhesion appendages. Such porous structures provide interconnected macroporous architectures for cell penetration and are desired in certain tissue engineering applications. With a tuned porosity the high surface area for cell growth and inter-connectivity for mass transfer of nutrients, fluids and gaseous exchange could be provided which are required to sustain cell proliferation. The structural porosity can be tuned with SUM printing and it should be explored in further studies. The confirmed cell attachment and proliferation demonstrated the absence of negative effects to cytocompatibility on the PCL arising due to SUM printing.

In the current study, cell migration into the samples was not observed (data not shown). This could be due to either insufficient cell culture time or too small pore size. To address this issue, the pore size could be increased by using a coarser powder combined with an optimized print feed rate. This would tailor the microstructure of the samples to support cell growth within the structure better.

During the printing, in addition to the thermal effects on PCL, cavitation occurs. Cavitation can form chemical species (*34*), some of which may be toxic to cells. Simulations of the acoustic field propagation revealed a negative pressure value of 17.4 MPa at the focus, corresponding to a mechanical index(MI) of 8.5 MPa MHz$^{-0.5}$ at an operating frequency of 4.17 MHz. High MI values indicate a high probability of cavitation, which typically occurs at MI values above 0.4, suggesting that cavitation occurred during printing. Despite this, the results of this work demonstrated high cytocompatibility of the printed samples, indicating that the cavitation process did not negatively affect cytocompatibility. These findings align with a study (*35*), which reported no toxic effects on cells from chemical species formed during cavitation in PCL diol. Reduction of MTT (3-(4,5-dimethylthiazol-2-yl)-2,5-diphenyl-2H-tetrazolium bromide) reagent to formazan product is a widely used colorimetric method to assess the metabolic activity of cells (*36*). Such a reaction is supposed to occur only in living cells, hence, the cell viability can be determined. Therefore, it was unexpected to observe a substantial background and formation of formazan crystals when the PCL powder alone or cell-free scaffolds were incubated with the MTT solution. FDA/PI staining was used as a mitigation strategy.

We presented the selective ultrasonic melting of PCL powder by using a focusing piezo transducer and a high-power piezo driver. The transducer was attached to a CNC machine to scan the focal point through a suspension of PCL and water. Prints were performed at an average acoustic power of 1.03 W. Scanning with feed rates from 25 mm min$^{-1}$ to 300 mm min$^{-1}$ resulted in solid porous objects.

      The proposed method of acoustic 3D printing outperforms the previous method with silicone rubber with regards to reusability, biodegradability, and potential for future applications. The presented method provides an alternative to methods such as FDM and binder jetting. With PCL established in tissue engineering and regenerative medicine, the proposed approach provides a new method for fabricating scaffolds. The outlined method of selective acoustic melting allows control of porosity by varying the feed rate which was confirmed with microstructure analysis using μCT and SEM. Using fluorescent microspheres it was confirmed that the structures possess an interconnected pore architecture, which can be tuned by altering the print parameters. The live-dead analysis and SEM imaging confirmed cytocompatibility of the printed structures, which is beneficial for biomedical applications. This also indicates that the fabrication method does not generate toxic by-products.

**Materials and Methods**
The Materials and Methods section should provide sufficient information to allow replication of the results. Begin with a section titled Experimental Design describing the objectives and design of the study as well as prespecified components.

In addition, include a section titled Statistical Analysis at the end that fully describes the statistical methods with enough detail to enable a knowledgeable reader with access to the original data to verify the results. The values for *N*, *P*, and the specific statistical test performed for each experiment should be included in the appropriate figure legend or main text.

All descriptions of materials and methods should be included after the Discussion. This section should be broken up by short subheadings. Under exceptional circumstances, when a particularly lengthy description is required, a portion of the Materials and Methods can be included in the Supplementary Materials.

**Ultrasonic 3D printer implementation**

The custom made SUM device comprises an electrical signal generator, an acoustic transducer, and a mechanical translation system. It is shown in **Figure 8**.
The electric signal was generated by a high-power piezo-driver which uses a single-ended MOSFET topology (*37*). This MOSFET produces a high-current signal which is transformed into a high-voltage signal by a transformer, that adjusts the output impedance to 50 W. A low-pass filter suppresses higher harmonic components at the output. The MOSFET is driven by a low-voltage logic level square-wave burst. For the experiments, the signal parameters were set at 80 cycles, with a frequency of 4.17 MHz, 50 % duty cycle, and a pulse-repetition-frequency of 1 kHz. This configuration produced sinusoidal bursts with 79 V amplitude. The average electric power consumption of the piezo driver during signal generation was 1.2 W, with a peak power of 62.4 W.
An ultrasonic focusing transducer was built from a piezo bowl (CTS Ferroperm, Denmark) with a radius of curvature of 15 mm, aperture of 19 mm, central hole diameter of 4 mm, and resonance frequency of 4.17 MHz. The transducer enclosure was 3D-printed, and the piezo bowl was glued to it with epoxy resin. A built-in matching circuit in the transducer adjusted the transducer's resistance to 50 W at the resonance frequency. The transducer was mounted on a Genmitsu 3018 3-axis CNC machine (SainSmart, USA) for positioning. The printing path was programmed using GCode that was loaded onto the CNC machine. The transducer's translation speed along the path (feed rate) was varied in from 25 mm min$^{-1}$ to 300 mm min$^{-1}$.

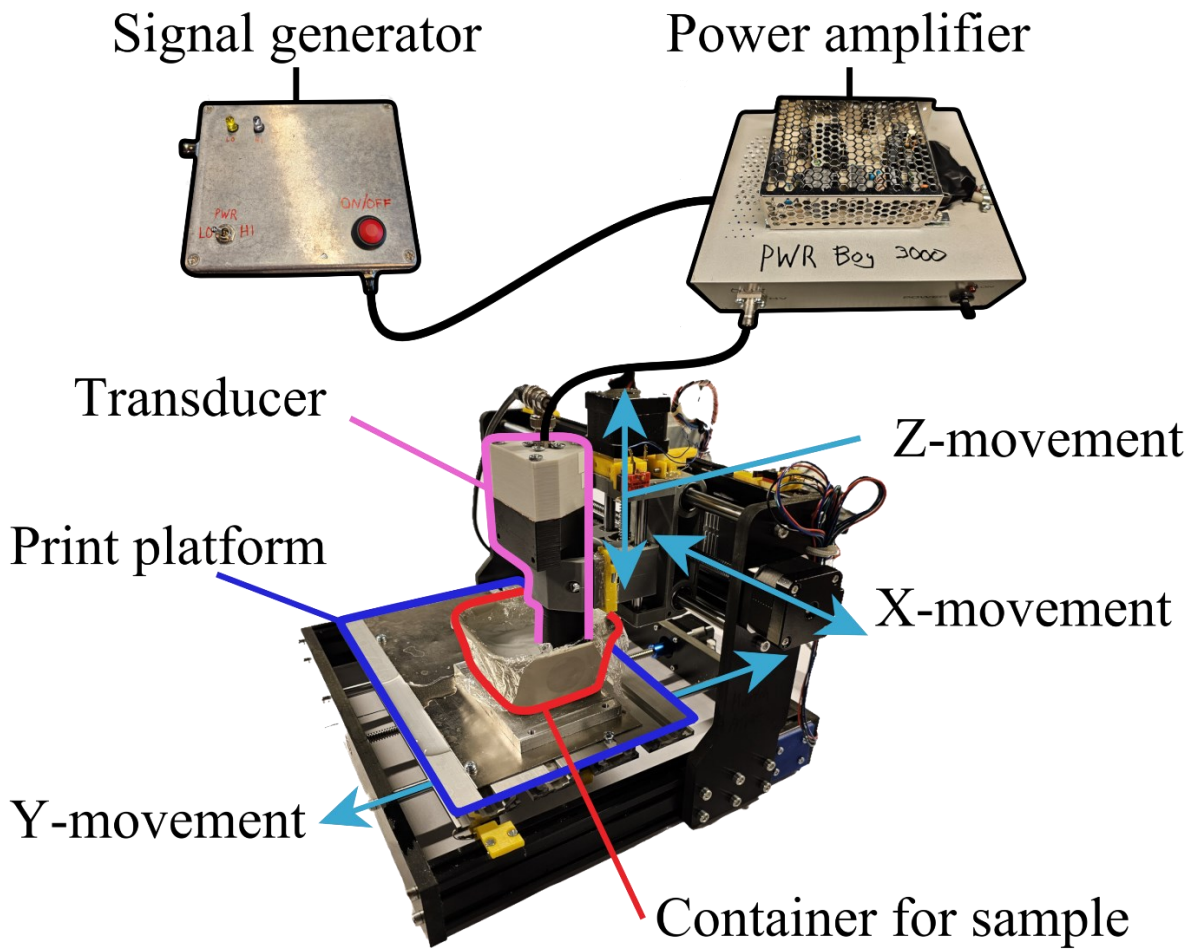

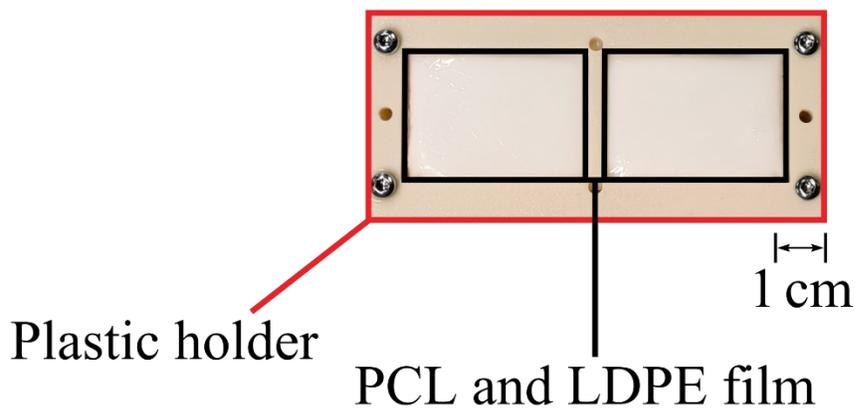

**Figure 8: Overview of the experimental setup.** The upper part illustrates the experimental setup, comprising the signal generator, power amplifier, and ultrasonic transducer mounted on the translation stage, allowing the transducer to move relative to the sample. The lower part shows the sample holder, which keeps the PCL-water suspension sample enclosed in a resealable LDPE bag.

**Printing material**

The printed material was prepared from medical grade PCL powder (Magerial Science, USA), with a number average molecular weight of 50 kg mol$^{-1}$ and a maximum particle size of 149 μm (mesh size 100). PCL is insoluble in water and has a melting point of around 60 °C (*38*). The powder was mixed with deionized, degassed water at a weight ratio of 3:2 to form a suspension, which was further degassed to remove trapped air. Excess water was then removed forming a 2:1 PCL-water suspension. The suspension was placed in an acoustically transparent resealable bag made of low-density polyethylene (LDPE) film with a thickness of 45 μm. The bag was fixed with a plastic holder and immersed into a container of degassed water which served as a sonic coupling medium between the sample and the acoustic transducer.

**Acoustic field characterization**

The acoustic field generated by the transducer for printing was characterized using a calibrated fiberoptical hydrophone (Onda, USA) in a plane transverse to the transducer axis positioned 5 mm from its surface. The scanning was performed in the nearfield as cavitation is present in the focal area. A 1 × 1 cm$^2$ area was scanned with a 200 μm step size, recording the full waveform at each point. Based on the scan data, the characteristics of the acoustic beam were calculated, including acoustic power and the vibrational velocity of the transducer surface (*39*).
The measured characteristics of the acoustic beam were used as input parameters for the *HIFU beam* software (*40*). The sound speed and attenuation coefficient of the PCL suspension, necessary for the simulation, were measured using the insertion loss method (*41*). The *HIFU beam* software was used to calculate the acoustic pressure, acoustic power, and thermal power in the focal area. These values allowed to estimate the sonication time required to melt the material as well as to determine the printing resolution.

**Micro computed tomography analysis**

Micro computed tomography (μCT) was used to investigate the micro-structure of the prints. Scans were conducted with the SkyScan 1272 (Bruker, USA) high-resolution 3D X-ray scanner with a voxel resolution of 7.5 μm and 15 μm. This method allows for non-destructive visualization of the printed 3D microstructure, as well as measurements of variations in material density and porosity. *Fiji ImageJ* software (*42*) was used to process CT data to calculate the porosity of the samples.

**Assessment of open porosity**

The Water intrusion method (*43*) was used to assess the open porosity of the material. Four disk shaped samples with a height of 4 mm and a diameter of 22.5 mm were printed at a feed rate of 180 mm min$^{-1}$. The samples were dried at room temperature for several days with their weights monitored using an analytical 410AM-FR balance (Precisa Gravimetrics AG, Switzerland) until a constant weight was achieved, at which point the samples were considered dry. Subsequently, they were immersed in deionized water at room temperature and degassed for 30 minutes to remove entrapped air. The samples were then taken out of the water and excess water was removed. The bulk porosity of the PCL scaffolds was calculated from the measured weights and known densities.
The sizes of the pore connections were evaluated using fluorescent microspheres (Polysciences, USA) with diameters of 1 μm and 10 μm. The microsphere suspension was diluted with water at a volume ratio of 1:200 and pipetted on top of each sample. The passed-through liquid was optically analyzed using the Axio Zoom.V16 fluorescent microscope (Zeiss, Germany) equipped with an HXP 120 V light source (LEJ, USA). The samples were cut in half to obtain cross-sectional images to evaluate the penetration of the microspheres. Fluorescent images were analyzed using ZEN 3.8 software (Zeiss, Germany).

**Live/dead imaging to assess cell viability**

A four-sector cell culture scaffold was designed, with each sector printed at different feed rates: 25, 75, 125, and 175 mm min$^{-1}$. Printing all feed rates in one sample minimized the potential impact of batch-to-batch variation in the printed material and reduced risks of microbiological contamination.

Murine NIH-3T3 fibroblasts (CRL-1658, ATCC) were cultured under standard conditions (37 °C, 5 % $CO_2$) in Dulbecco's Modified Eagle Medium (DMEM, high glucose, GlutaMAX™ Supplement, pyruvate; Gibco™, Thermo Fisher Scientific Inc., Waltham, USA) supplemented with 10 % calf serum (iron supplemented, US origin, Hy-Clone™), 1 % 100 IUmL$^{-1}$ penicillin and 100 μgmL$^{-1}$ streptomycin (Gibco™). The cells were passaged 2-3 times a week at the subculture ratio of 1:10 - 1:15 using TrypLE™ Express Enzyme (Gibco™).

The printed scaffolds were sterilized with 70 % ethanol for 1 hour, washed 3 times with Dulbecco's phosphate buffered saline (DPBS, no calcium, no magnesium, Gibco™), and soaked in the calf serum for 1 hour. Then the scaffolds were placed in the complete DMEM medium supplemented with 0.25 μg mL$^{-1}$ amphotericin b and 10 μg mL$^{-1}$ gentamicin (Gibco™) to inhibit potential fungal or bacterial contamination and left in the incubator overnight.

The next day, $1.5 \times 10^5$ NIH-3T3 cells were seeded on the scaffolds and cultured in a 12-well plate. The cells cultured in the well without a scaffold were used as a positive control to verify the viability of the cells. After three days, the cell-containing scaffolds were stained with 8 μg mL$^{-1}$ fluorescein diacetate (FDA, BLD Pharmatech Ltd., Shanghai, China) and 20 μg mL$^{-1}$ propidium iodide (PI, BLD Pharmatech Ltd.) dissolved in DMEM medium supplemented with 2 % calf serum. The staining was done at room temperature in the dark for 5 minutes followed by two washings with DPBS. The live and dead cells were imaged with a fluorescence microscope (ZEISS Axio Zoom.V16, Germany). The z-stack images were analyzed with the ZEISS ZEN 3.8 software.

Cell viability was calculated based on the live/dead images according to a published protocol (*44*) for automatic quantification of live and dead cells using Fiji-ImageJ.

**SEM images of cell-containing scaffolds**

After verifying the presence of the cells on the scaffolds by live/dead imaging, the same cell-containing scaffolds were fixed with the paraformaldehyde solution (4 % w/v) in 0.2 mol L$^{-1}$ sodium cacodylate buffer pH7.2 for four hours at 4 °C. The samples were rinsed with PBS followed by deionized water. The samples were then dehydrated via stepwise sequential ethanol gradient method, starting from 30 % v/v to 100 % v/v with a hold time of 30 min at each step. The samples finally dried by placing in a desiccator for 24 h to remove traces of moisture. Before imaging, the samples were sputter coated with 5 nm of platinum. The SEM imaging was performed using FEI Quanta-250 (FEI, Hillsboro, USA). The control 3D-scaffolds were processed in similar manner for SEM imaging to analyze surface topography, however they were not seeded with cells.

**Acknowledgments**

We thank Heikki Suhonen for μCT processing of the samples. We also thank the SEM core facility supported by University of Helsinki and Biocenter Finland. This research was supported by the HiLIFE Proof of Concept funding provided by University of Helsinki and Helsinki Innovation Services Incentive funds. Funding was provided by the Research Council of Finland grants 347459, 349200. The Helsinki University Library funded open access.

**Author contributions:**
    Conceptualization: MW, DN, JH, ET, RL, AS
    Methodology: MW, DN, ET, RL
    Software: MK
    Formal analysis: DN, MK, JJ, RL
    Investigation: DN, MK (printing), JJ (printing, microscopy), AN (μCT), ET (cell studies, microscopy), AT (SEM analysis).
    Resources: JH, RL, AS
    Visualization: MK, ET
    Supervision: MW, PK, RL, EH, AS
    Writing—original draft: MW, DN, MK, JJ, ET, AT
    Writing—review & editing: MW, DN, MK, JH, JJ, AN, ET, AT, PK, RL, EH, AS
    Project administration: EH, AS
    Funding acquisition: MW, AT, PK, RL, EH, AS

**Competing interests:** University of Helsinki has filed a patent application under the number FI 20245641 related to the 3D printing method. The inventors are MW, AS, Arto Klami, RL, DN, JH, MK, JJ, EH, Anton Nolvi, and Denys Iablonskyi. All other authors declare they have no competing interests.

**Data and materials availability:** Data is available on request from the corresponding author.

## Supplementary Materials

Please refer to the supplementary material provided for Figure S1 to Figure S6.

# Supplementary Materials for

- **Fabrication of Poly (ε-Caprolactone) 3D scaffolds with controllable porosity using ultrasound**

Martin Weber *et al.*

*Corresponding author. Email: martin.weber@helsinki.fi

**This PDF file includes:**

Supplementary Text
Figs. S1 to S6

# Supplementary Text
## Serum choice influenced the cell survival

Additional images to complement Fig.7. The fractions of live cells (viability) were calculated using Fiji-ImageJ according to the protocol (*44*). 10 % calf serum was used to supplement the complete DMEM medium.

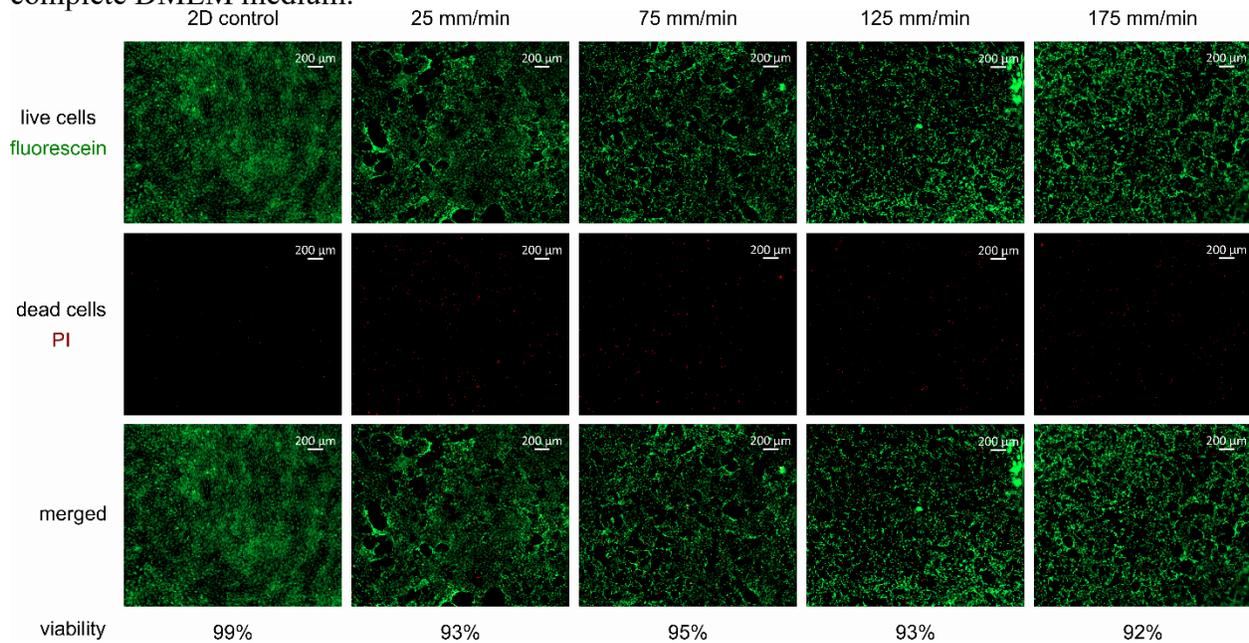

**Figure S1 Viability and distribution of NIH-3T3 fibroblasts cultured on the ultrasound printed four-sector scaffold.** The scaffold was sterilized before seeding the cells. NIH-3T3 fibroblasts were cultured for three days on the PCL scaffold printed with 25, 75, 125, and 175 mm min$^{-1}$ feed rates followed by the FDA/PI staining. Live cells were detected with the green fluorescent signal (fluorescein), and dead cells with the red signal (PI). Cells cultured without the scaffold (2D control) were used as a positive control.

In the following experiment, NIH-3T3 fibroblasts were cultured in the complete DMEM supplemented with 10 % Fetal Bovine Serum (FBS Premium, Gibco™). Here, the cells cultured on the four-sector scaffold could not survive, whereas control cells showed excellent viability.

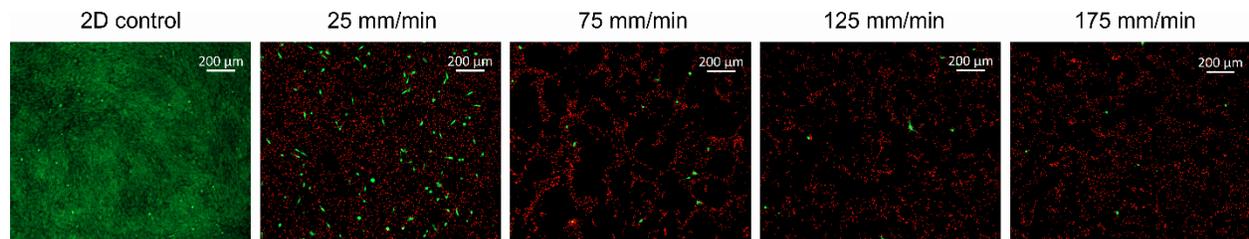

**Figure S2 Viability of NIH-3T3 fibroblasts cultured in complete DMEM medium supplemented with FBS on the ultrasound printed four-sector scaffold.** The scaffold was sterilized before seeding the cells. NIH-3T3 fibroblasts were cultured for three days on the PCL scaffold printed with 25, 75, 125, and 175 mm min$^{-1}$ feed rates followed by the live (green) / dead (red) staining. The cells cultured without the scaffold (2D control) were used as a positive control.

## NaOH treatment

Sodium hydroxide (NaOH) etching modifies a PCL surface by increasing its hydrophilicity and roughness. Treating the PCL scaffold with NaOH enhanced cell attachment to it (*22,41,45*). However, in the case of ultrasound-printed PCL scaffolds, there was no such evidence under studied conditions. The NIH-3T3 fibroblasts cultured in the complete DMEM medium supplemented with FBS did not attach to the NaOH treated PCL scaffold. On the other hand, the cells cultured in the presence of calf serum showed comparable adhesion, viability and distribution as observed for the untreated scaffold. SEM analysis of the cell-seeded scaffolds confirmed the presence of cells growing on the scaffold surfaces. On the NaOH treated surfaces, the cells grew either as monolayers on the flat surfaces or formed cellular appendages providing 3D extended morphology. However, with SEM analysis we were unable to see any significant difference in the cell adhesion and proliferation between the non-treated and NaOH treated scaffolds. This demonstrates that the NaOH treatment of the scaffolds caused no negative effect on cell adhesion and viability.

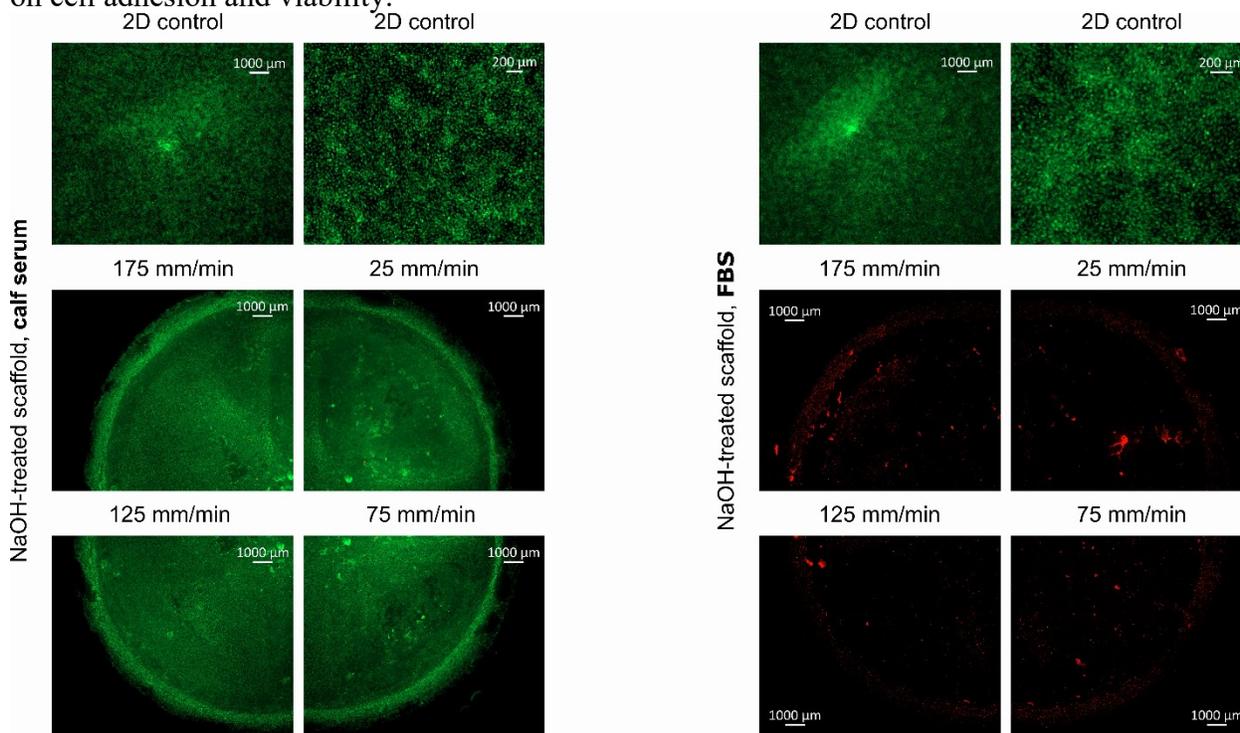

**Figure S3 The viability of NIH-3T3 fibroblasts cultured on the NaOH treated PCL scaffolds in the presence of calf serum or FBS.** The four-sector PCL scaffolds were immersed in the 5 mol L$^{-1}$ NaOH solution for 30 min and, subsequently, washed several times with DI water until the pH became neutral. Both scaffolds were sterilized as described in Material and Methods, but one scaffold was coated with calf serum, and the other - with FBS. The next day, NIH-3T3 fibroblasts were seeded onto the scaffolds and cultured either in calf serum or FBS, respectively. The live (green) / dead (red) imaging was performed three days later. The representative images are shown for the different sectors printed with 25, 75, 125, and 175 mm min$^{-1}$ feed rates. The cells cultured without the scaffold (2D control) were used as a positive control**.**

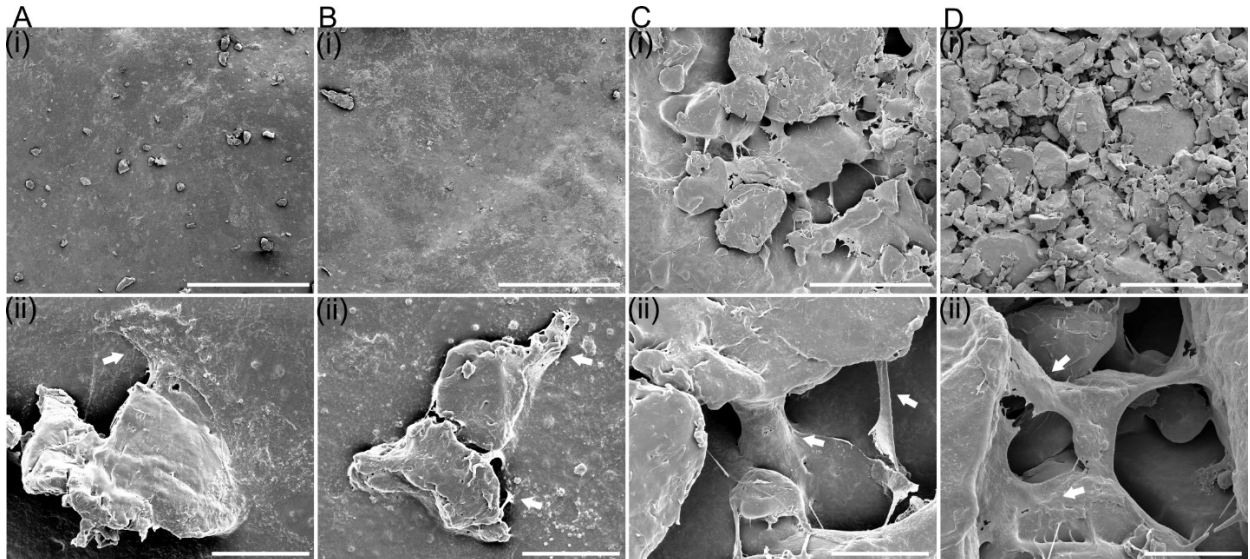

**Figure 4: SEM image of cells cultured on NaOH treated PCL scaffolds.** Images showing surface topography of 3D scaffolds printed at different feed rates. (A) 25 mm min$^{-1}$, (B) 75mmmin−1, (C) 125 mm min$^{-1}$, and (D) 175 mm min$^{-1}$. (i), (ii) showing images of cells (white arrows) growing on the respective scaffold surface at different magnifications. Scale bar- (i)500 μm, (ii)20 μm.

MTT generated background signal in cell-free scaffolds

The MTT (3-(4,5-dimethylthiazol-2-yl)-2,5-diphenyltetrazolium bromide) assay is used to assess cell proliferation and viability (*36*). This method was applied to determine the cytotoxicity of PCL scaffolds made by electrospinning (*46*). Therefore, it was unexpected to detect a substantial background of 6 – 16 % generated by the ultrasound-printed PCL scaffolds in case no cells were seeded onto them. Hence, we used another method to assess cell viability. Live and dead cells were visualized by FDA/PI staining, which generated little background.

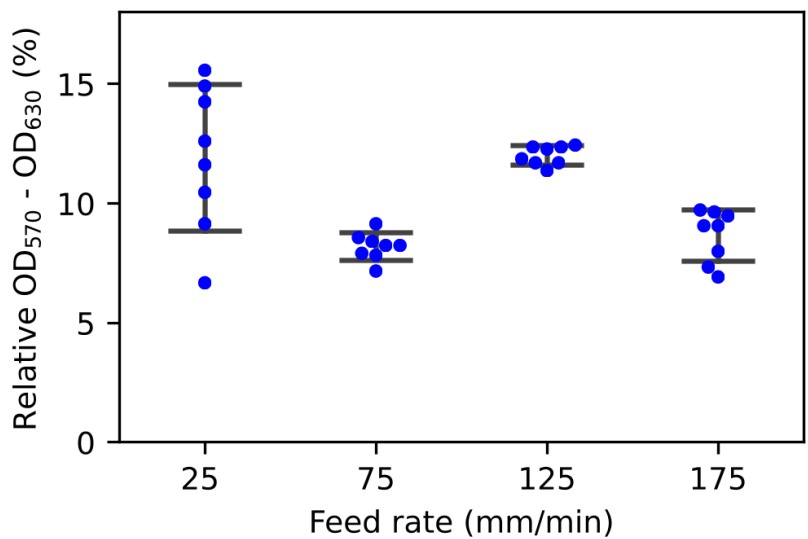

**Figure S5: Relative light absorbance in MTT assay.** Four PCL scaffolds were printed with different feed rates of 25, 75, 125, and 175 mm min$^{-1}$ and handled as described in Material and Methods with one exception: no cells were seeded onto those scaffolds. After three days in culture, the scaffolds were incubated with 0.5 mg/mL MTT

dissolved in the DMEM medium. During this time the scaffolds changed color from white to violet, which indicates that MTT was reduced to formazan. The MTT solution was exchanged for dimethyl sulfoxide (DMSO) to dissolve the formazan product. The light absorbance of each sample was measured at 530 nm (signal) and 630 nm (reference). The cells cultured without the scaffold (2D control) were used to normalize the data and set to 100 %. Data are presented as mean ± standard deviation. Each dot is a technical replicate, n = 8.

Printed sample for microstructure analysis of different feed rates

A ladder-like structure was printed study the microstructure created by different feed rates, Fig.4. All feed rates were done into a single print as seen in Fig.S6 to ensure that they are comparable. The order of different feed rates was carefully chosen so that the slower and more stable feed rates were at the edges providing additional support.

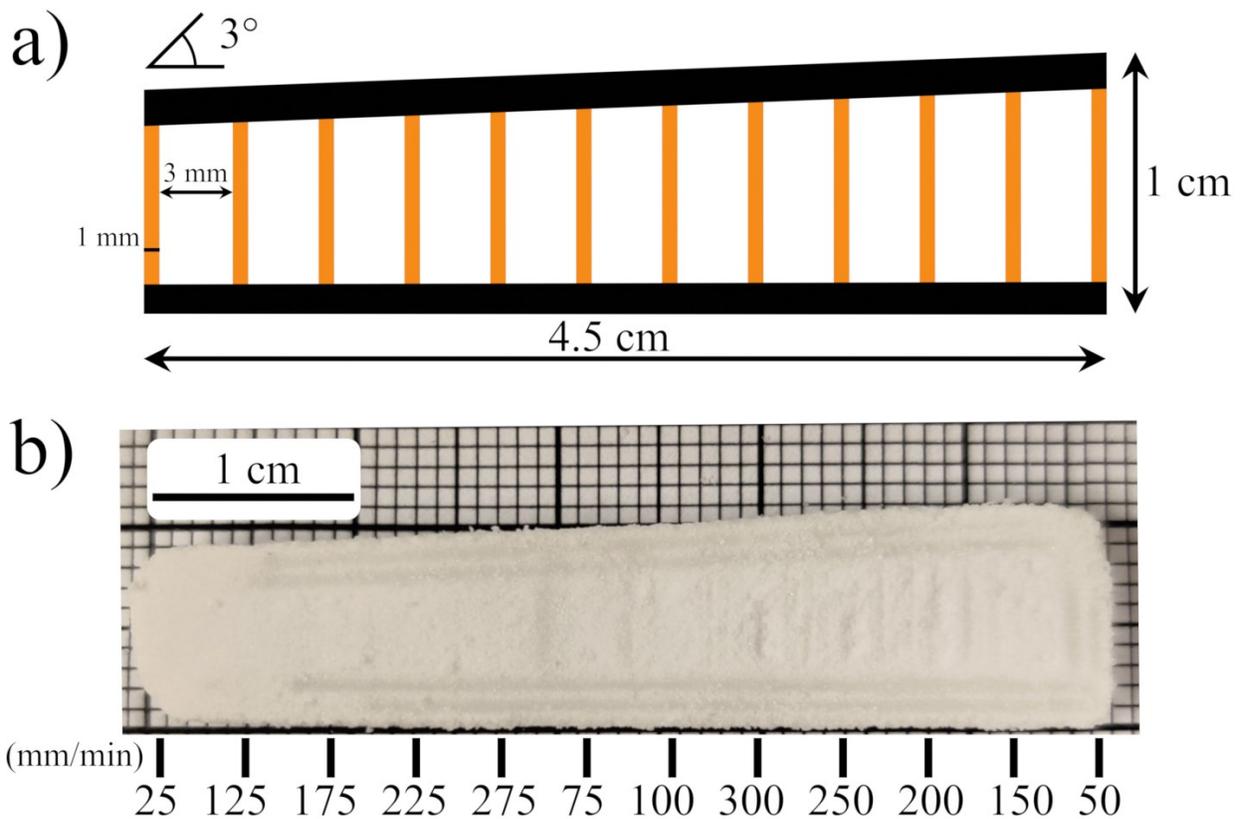

**Figure S6: The schematic of the ladder structure print (a) and the printed result (b).** The ladder structure print features speeds from 25 mm min$^{-1}$ to 300 mm min$^{-1}$ with steps of 25 mm min$^{-1}$. Each feed rate was printed along the orange vertical lines as shown in part a. Every vertical line consists of two printed paths with a separation of 1mm and each individual feed rate line is separated by 3mm. The specific feed rates are faintly visible in image b and loose powder is filling the empty gaps.